# Metal-free magnetism in chemically doped covalent organic frameworks

Hongde Yu and Dong Wang[*]

MOE Key Laboratory of Organic OptoElectronics and Molecular Engineering, Department of Chemistry, Tsinghua University, Beijing 100084, PR China

*Supporting Information*

**ABSTRACT:** Organic and molecule-based magnets are not easily attainable, as introduction of stable paramagnetic centers to pure organic systems is particularly challenging. Crystalline covalent organic frameworks (COFs) with high designability and chemical diversity constitute ideal platforms to access intriguing magnetic phenomena of organic materials. In this work, we proposed a general approach to attain unpaired electron spin and metal-free magnetism in narrow-band COFs by chemical doping. By using density functional theory calculations, we found that dopants with energy-matched frontier orbitals to COFs not only inject charges but also further localize them through orbital hybridization and formation of supramolecular charge-transfer complex. The localized electronic states ensure that stable paramagnetic centers can be introduced to nonmagnetic COFs. Based on these discoveries, we designed two new COFs with narrow valence bands, which show prospective magnetism after doping with iodine. Further, we unraveled the magnetic anisotropy in two-dimensional COFs and demonstrated that both spin-conduction and magnetic interactions can be effectively modulated by manipulating the building blocks of COFs. Our work highlights a practical route to attain magnetism in COFs and other organic materials, which show great potential for applications in organic spintronic devices.

## INTRODUCTION

Crystalline two-dimensional (2D) polymers with columnar π-arrays and inherent pores, which are also known as covalent organic frameworks (COFs), can integrate organic building blocks into periodic structures via reticular chemical reactions.[1-2] Since their first synthesis in 2005,[3] these emergent materials have attracted considerable interests for diverse building blocks and designable linkage topologies. These attributes make COFs promising platforms to explore the interplay between charge carriers, phonons, and excitons, and display immense potential in a large variety of applications, such as gas storage[4] and separation,[5-6] catalysis,[7-8] energy conversion[9-10] and ion conduction.[11-12] Beyond these, magnetic ordering is a new and important property that can be associated with COFs, as manifested by the observation of ferromagnetic phase transition in a fully π–conjugated, iodine doped $sp^2$c-COF.[13]

Organic and molecule-based magnets whose unpaired electron spins reside in p-orbitals are a new and growing class of magnetic materials exhibiting ferromagnetic or other types of magnetic behavior.[14-15] They are light-weight, chemically diverse, transparent, and easy to process. However, most organic materials are lack of unpaired electrons, except for radicals of small molecules or polymers, which are highly reactive and prone to redox reactions. The instable nature of radicals severely obstructs fabrication of organic magnets. In organic electronics, either chemical or electrochemical doping has been widely applied to control the electrical conductivity of conjugated, small-molecule or polymeric semiconductors. In the process of doping, not only charges are injected, but also stable unpaired electron spins can be introduced to the electronic structure.

Although some researches[16-17] attempted to attribute the magnetism in COFs to "flat bands" arising from certain symmetry of lattices, how chemical compositions of COFs impact their band structure and the role of chemical dopants are beyond discussion. So, the approach to introduce magnetism to nonmagnetic COFs by chemical doping is not yet established. In this work, we proposed a general method to generate paramagnetic centers and attain metal-free magnetism in chemically doped COFs, which requires both localized electronic states of COFs and matched energy levels between COFs and dopants. Based on this approach, we designed several new COFs which show prospective magnetism after doping. The relationships between chemical compositions of COFs and magnetic properties are investigated, and the effects of a variety of chemical dopants are also explored. Furthermore, we unraveled anisotropic magnetisms in these COFs and demonstrated that both spin-conducting behaviors of metal and half-metal and magnetic interactions can be modulated by engineering building blocks. These findings offer immense opportunities for chemists to fabricate COFs with intriguing magnetic behaviors, and will tremendously boost their applications as spin-crossover,[18-19] spin liquid[20] and other spintronic materials.[21]

## RESULTS AND DISCUSSION

### Introducing paramagnetic centers to non-radical COFs.

Two basic steps to attain magnetism in materials are, (1) introducing paramagnetic centers, (2) realizing magnetic interactions between them.[22] Unlike inorganic magnetic materials with half-filling d or f orbitals in transition metals, absence of unpaired electron spins in organic materials hinders the formation of paramagnetic centers. One of solutions is to utilize radical compounds.[23] However, owing to their highly reactive nature, radical materials are sensitive to oxygen or solvent, which obstructs their practical applications. Although many researchers have observed paramagnetic signals with electron spin resonance in organic materials upon chemical doping,[24] these magnetic signals undergo rapid annihilation with the increasing dopant concentration due to strong electronic couplings and large hopping integrals in π–conjugated polymers or molecular crystals. To avoid electron pairings in organic materials, a well-known strategy is to employ the so-called "flat bands", namely localized electronic states.[25] According to the

Stoner criterion,[26] ferromagnetic phenomenon can be attributed to sufficiently small bandwidth in iterative systems with partially filled energy bands. However, except for some special lattice symmetries such as the Kagome lattice,[27-28] flat bands in pristine COFs are not easily attainable. In the following, we propose a strategy to facilitate organic magnetism in chemically doped COFs with narrow energy bands. We manifest that, by careful design dopants not only inject charges to the host material via charge-transfer interactions, but also further localize the charges, preventing them from pairing and stabilizing paramagnetic centers even at high dopant concentration. In this situation, paramagnetic centers will not annihilate with charge injection, and magnetic ordering in COFs is likely to be achieved.

**Facilitation of narrow band in pristine COFs by rational design.**

COFs are constructed by nodes, edges and linkages between them. In order to attain localized electronic states and narrow energy bands in COFs, we propose three strategies: (1) Enlarging the energy mismatch between frontier orbitals of building blocks; (2) Increasing the torsion angle between building blocks utilizing the stereo-hindrance effect; (3) Applying non-conjugated linkages such as boronate esters. Based on these strategies, four COFs with pyrene as nodes and linear edges and linkages have been designed, and they are PPy-CC-Ph, PPy-CC-Naph, PPy-BO-Ph and PPy-CC-4FPh as sketched in Scheme 1. Herein, PPy represents the pyrene node and it is short for tetrakis(4-formylphenyl)phenyl; Ph, Naph, and 4FPh represent the edges, which are short for phenyl, naphthyl and 2,3,5,6-tetrafluorophenyl; CC and BO are short for C=C and boronate ester linkages.

PPy-CC-Ph, also known as $sp^2c$-COF, is C=C-linked with fully π-conjugated structure. The ferromagnetic phase transition of iodine doped $sp^2c$-COF was first reported by Jiang and co-workers.[13] The band structure and charge density distribution of pristine PPy-CC-Ph are shown in Figure 1. It is a semiconductor with indirect band gap of 1.16 eV calculated by the PBE-D3 method (Figure 1a) and 1.7 eV calculated by the HSE method. It consists of a narrow valence band with the bandwidth of 0.23 eV and a dispersive conduction band with the bandwidth of 0.58 eV. The charge density at the valence band maximum (VBM) localizes mainly on the pyrene node and exhibits the π–conjugated bonding feature, while at the conduction band minimum (CBM) it is delocalized over the entire structure and shows antibonding character (Figure 1b). As illustrated by the energy level diagrams in Figure 2a and Figure S1, the HOMO energies of its PPy node and Ph edge differ by over 2.6 eV. As a result of the large energy mismatch between them and much higher energy of PPy, the HOMO of PPy-CC-Ph is localized on its node and a narrow valance band is identified in the periodic structure. In contrast, the LUMO energies of node and edge match well with only slight difference of 0.1 eV, which gives rise to the wide conduction band of PPy-CC-Ph. Moreover, as illuminated by the partial density of states (pDOS) in Figure S2, both CBM and VBM are constituted dominantly by C-$2p_z$ orbitals, which confirms the formation of extended π-conjugation.

PPy-CC-Naph is also linked by -C=C- and possesses fully π-conjugated structure. However, replacement of phenyl with a larger linear edge of naphthyl increases the stereo-hindrance, which leads to an enlarged torsion angle of 39.6° between node and edge, in contrast to 22.8° in PPy-CC-Ph (Figure 2b). The band structure of PPy-CC-Naph (Figure S3) shares the same feature as PPy-CC-Ph with the charge density of VBM (Figure S4) also localized on the node. Due to the stereo-hindrance, both valence band and conduction band become narrower with the bandwidth of 0.14 eV and 0.38 eV respectively. To increase both energy mismatch and structural distortion, PPy-CC-4FPh is constructed by replacing the Ph edge in PPy-CC-Ph with 4FPh. 4FPh is a widely used building block in polymer chemistry,[29] and its frontier orbital energies are substantially lower than those of Ph due to the strong electron-withdrawing effect of F, resulting in even larger energy mismatch between HOMOs of node and edge. Since the atomic radius of F is larger than H, the stereo-hindrance in PPy-CC-4FPh is increased and the torsion angle is enlarged to 34.3° (Figure 2b). Its band structure is similar to the other two –C=C– linked COFs (Figure S5), and the charge density at VBM is localized on the node (Figure S6). The valence bandwidth in PPy-CC-4FPh is 0.14 eV, smaller than 0.23 eV in PPy-CC-Ph. In PPy-BO-Ph, the pyrene node and Ph edge are linked by boronate ester instead of -C=C-. Boronate ester is the first and most commonly used linkage in synthesized COFs, and it is well recognized for breaking the in-plane conjugation.[3, 30] As a result, both valence band and conduction band have been narrowed (Figure S7), and charge densities at VBM and CBM are highly localized on the node (Figure S8). Therefore, all the four COFs mentioned above consist of a narrow valence band which is mainly localized on the PPy node.

For comparison, we also designed two COFs with wider valence bands and more dispersive electronic structure. They are PPy-CC-1T and PPy-CN-Ph (Scheme 1). Here 1T is short for thienyl and CN is short for the imine linkage, -C=N-. PPy-CC-1T is another -C=C- linked COF with the thienyl edge, which is a popular building block in both COFs and conjugated polymers.[31] PPy-CN-Ph is constructed by the pyrene node and Ph edge similar to PPy-CC-Ph, but linked by –C=N–, which is another extensively explored type of linkage in experiments and also capable of forming fully π-conjugated structure.[32-33] Owing to the better energy level matching between the pyrene node and 1T edge in PPy-CC-1T (Figure 2a) and the electron-rich character of –C=N– linkage in PPy-CN-Ph, their valence bandwidths are 0.42 eV and 0.58 eV respectively, almost twice as large as PPy-CC-Ph, which implies much more delocalized electronic structures (Figures S9-12).

All these COFs share the same Lieb lattice, and are lack of flat bands near the Fermi level. Nevertheless, we demonstrate that their bandwidths can be tuned to a large extent via chemical design of building blocks. The narrow bands are needed to attain magnetism by chemical doping, while the wide bands are desired for high-performance charge transport. Based on the three strategies proposed, we have achieved charge localization in COFs. These design strategies, including energy level matching, stereo-hindrance effect and choice of linkages, are robust and we will show below that not only electronic structures but also magnetic structures can be modulated by engineering chemically diverse building blocks of COFs.

**Attaining spin-polarized COFs by chemical doping.**

To introduce paramagnetic centers to nonmagnetic COFs, chemical dopants must be added to inject unpaired electron spins. However, it was often observed that in organic π-conjugated polymers the spin signal could be generated at the start of doping but the magnetic moment quickly diminished with the increase of doping level, which mainly arises from the strong electronic coupling in conjugated systems that tends to pair the spin-polarized electrons. On the charge transfer between dopants and organic semiconductors, two mechanisms have been proposed depending on the redox strength of dopants, which is either the formation of ionic pair or supramolecular complex.[34-35] If a p-dopant is strong oxidant, which means the energy of its lowest unoccupied molecular orbital (LUMO) or singly occupied molecular orbital (SOMO) is much lower than the highest occupied molecular orbital (HOMO) of COFs, and the coupling between these orbitals is weak, an integral charge will be transferred from COFs to the dopant, and ionic pair is formed in which the hole is completely located on COFs (Figure 3a). In this case, doping will not cause sufficient energy split for opposite spins, due to the strong electronic coupling in COFs even they have narrow valence band. Otherwise, if a p-dopant is weak and its frontier orbital energy is close to that of COFs, orbital hybridization will occur and supramolecular complex is formed in which only fractional charge is transferred (Figure 3a). As a result, both charges and spins are shared by COFs and the dopant. In the following, we demonstrate that formation of supramolecular complex and orbital hybridation with the dopant will further flatten the valence band of COFs due to the localized nature of dopant orbitals. By chemical doping with energy-matched dopants, paramagnetic centers will be introduced to nonmagnetic COFs.

We take PPy-CC-Ph as an example to illustrate the effect of chemical doping at high dopant concentration. Due to the porous nature of COFs, dopants are deposited into pores every other layer as shown in Figure 3b and Figure S13, and the ratio of pyrene nodes to dopants is 1:1, which will give one spin per pyrene node, close to the spin concentration of 0.7 per pyrene unit observed in the experiment. Various dopants of either p-type or n- type have been added to PPy-CC-Ph to investigate the effect of chemical doping. when applying n-dopants of K and Na, integral charge transfer occurs and the electron is injected to the conduction band of COFs. However, due to the strong electronic coupling and large hopping integral in the dispersive conduction band of PPy-CC-Ph, energy split for opposite spins is not observed and paramagnetic centers are generated (Figure S14 and Table 1). The absence of magnetism in heavily doped PPy-CC-Ph with K and Na is attributed to delocalization of the electronic states in pristine COFs as well as formation of ionic pair upon doping.

For p-doping, several dopants including I, Br, Cl and $CH_3SO_3$ with varying SOMO energies have been applied. As the SOMO energy decreases (I > Br > Cl > $CH_3SO_3$) and electron affinity of the dopants increases (Figure 3c and Figure S15), the energy mismatch between COFs and the dopants enlarges. Consequently, the charge transferred changes from fractional to nearly integral (Table 1), and the doping mechanism transits from the formation of supramolecular complex to ionic pair. Both band structure and density of states (DOS) of I-doped PPy-CC-Ph show that two spin-polarized subbands are vacant due to the charge transfer from COFs to the dopant. And there exist appreciable energy splits for opposite spins in I, Br and Cl-doped COFs but only tiny split in $CH_3SO_3$ doped one (Figure 3d, Figure 3e and Figure S16), especially for the valence bands around the Fermi level, which are responsible for the generation of paramagnetic centers. In the case of iodine doping, both $sp^2$-C orbitals of COFs and I orbitals contribute to the electronic states in the spin-polarized hole bands (Figure 3e), and spin density is located on both the pyrene node of COFs and iodine dopant (Figure 3f). Moreover, the hole bandwidth (0.12 eV) of I-doped COF is much smaller than the valence bandwidth of pristine COF (0.23 eV), implying that the formation of supramolecular complex further flattens the band and thus benefit generation of paramagnetic centers. With the SOMO energy decreasing from I to Br and Cl, the dopant contribution to the hole band reduces, as a result the hole bandwidth increases, while spin density gradually diminishes though it is still shared by COFs and the dopant (Figure S16 and Table 1). When it comes to $CH_3SO_3$ with the deepest SOMO, whose anion often serves as counter-ion in the experiment,[36] the hole band arises solely from COFs and its bandwidth of 0.19 eV almost recovers 0.23 eV of pristine COFs (Figure 3e). We observed tiny energy split for opposite spins, ~ 6 meV, and negligible spin density (Figure 3f and Figure S16). Correspondingly, the magnetic moment of doped COFs decreases from about one $\mu_B$ per node to 0.03 $\mu_B$ per node and the energy difference between spin-polarized and spin-unpolarized states, $\Delta E_P$, also reduces, indicating diminished paramagnetic centers (Table 1).

The Hubbard model is widely used to discuss the magnetism in materials, in which two terms with competitive effects are considered, namely, the on-site Coulomb repulsion between two electrons or holes with opposite spins, denoted as $U$, and the inter-site electronic coupling or hopping integral between nearest-neighbors, denoted as $t$. Notably, $t$ accounts for the formation of spin-unpolarized energy band and it is related to the bandwidth via $W = 2zt$ ($z$ is the coordination number of each site). Moreover, the $U/t$ ratio is responsible for the stability of paramagnetic centers and phase transitions in magnetic materials such as the metal to Mott insulator transition. Sufficiently large $U/t$ ratio will stabilize paramagnetic centers and result in a variety of interesting magnetic correlations including spontaneous ferromagnetic or antiferromagnetic order, spin liquid and so on.[20] In the mean-field approximation, the well-known Stoner criterion can be derived from the Hubbard model, which indicates that metals become magnetic when $UD_F > 1$ or $U/W > 1$, here $D_F$ is the DOS at the Fermi level, and is approximately $1/W$. Therefore, spontaneous magnetism in itinerate systems could only be achieved when the DOS at the Fermi level is sufficiently large, or in other words, when there exist flat bands.

We have derived $U$ and $t$ parameters from the band structures of chemically doped COFs. $U$ is taken as the energy split for opposite spins around the Fermi level and $t$ is estimated from the bandwidth, $W$. Similar method[27, 37] was previously applied to obtain $U$ and $t$ in radical COFs. As the energy mismatch between COFs and the dopant increases with the dopant going from I to Br and Cl, the $U/W$ or $U/t$ ratio gradually decreases, because the bandwidth $W$ or hopping integral $t$ increases while $U$ decreases (Table 1, Table S1 and Figure S16). Furthermore, when COFs are doped with $CH_3SO_3$, $U$ and $U/W$ drop noticeably, hence stable paramagnetic centers are not generated. Our first-principles results of chemically doped COFs are in accordance with the Stoner criterion, i.e., the flatter band and higher DOS around the Fermi level give rise to larger magnetic moment. A promising strategy to facilitate flat band and introduce paramagnetic centers to COFs with narrow energy band, is to apply proper chemical dopants whose frontier orbital energy matches that of COFs to enable orbital hybridization.

## Modulation of spin-conduction and magnetic interactions in doped COFs.

As shown above, we have attained four COFs with narrow valence band, PPy-CC-Ph, PPy-CC-Naph, PPy-BO-Ph and PPy-CC-4FPh, whose VBMs are mostly localized on the pyrene node. Similar to PPy-CC-Ph, doping of PPy-CC-Naph, PPy-CC-4FPh and PPy-BO-Ph with iodine will also lead to the formation of supramolecular charge-transfer complex and introduction of paramagnetic centers. As confirmed by band structure and DOS, the energy of opposing spins is split and spin-polarized holes are injected to COFs (Figure 4a and Figure 4b). Both charge and spin densities are located on the pyrene node and the iodine dopant (Figure S17 and Figure S18). Spin-polarized electronic structures are more stable than spin-unpolarized ones, with energies at least 40 meV lower (Table 2), indicating formation of robust paramagnetic centers on these I-doped COFs. In contrast, valence bands of pristine PPy-CC-1T and PPy-CN-Ph are highly dispersive with the bandwidth twice as large (Table 2) and charge densities delocalized over the entire framework (Figure S10 and Figure S12). As a result, the energy splits of opposing spins around the Fermi level reduce (Figure S19 and Figure S20), and the magnetic moments decrease (Table 2), implying reduced stability of paramagnetic centers.

Notably, PPy-CC-Ph and PPy-BO-Ph are metallic and partially filled for both up and down spins (Figure 3d and Figure S18), while PPy-CC-Naph and PPy-CC-4FPh are half-metallic where only bands for down spin are partially filled and there exist large gaps in DOS for up spin (Figure 4a and Figure 4b). Due to the energy gap asymmetry for different spin channels, only carriers with down spin display conducting behavior while the other spin channel is insulating, so spin-polarized current can be obtained in half-metallic magnetic COFs. These findings indicate that both metallic and half-metallic structures as sketched in Figure 4c could be achieved in chemically doped COFs by subtle modification of building blocks, which holds great promise for exploring spin-based electronics with COFs.

Below, we investigate magnetic interactions in iodine-doped COFs, especially the energy difference between antiferromagnetic (AFM) and ferromagnetic (FM) phases, which is directly related to the spin exchange parameter $J_{ij}$ ($J = E_{AFM} - E_{FM}$) in the context of Heisenberg spin Hamiltonian defined as $H_s = -\sum_{i,j} J_{ij} \vec{S}_i \vec{S}_j$.[38] 2D COFs with layered structures are covalently bonded within the layer and π-π stacked between layers, so magnetic interactions in them are likely to be anisotropic. In general, the spin exchange arises from two competitive interactions, i.e., potential and kinetic exchanges, and it can written as $J = K - 4t^2/U$.[39-40] In this expression, potential exchange, denoted as $K$, originates from direct exchange between parallel spins on neighboring sites. This interaction is short-ranged and is effective only when hopping integral $t$ approaches zero, which will lead to the parallel spin alignment according to Hund's rule and therefore account for the ferromagnetic interaction. And kinetic exchange, denoted as $-4t^2/U$, is derived from second-order perturbation theory and it originates from hopping of spin-polarized electron between neighboring sites. The kinetic exchange could be long-ranged mediated by overlapped orbitals and it results in the antiparallel spin alignment and antiferromagnetic interaction as a consequence of Pauli's principle.

As indicated by spin density distributions of I-doped PPy-CC-Ph (Figure S21), interlayer magnetic interaction between neighboring π-stacks of pyrene is dominated by through-space potential exchange $K$, due to vanishingly small hopping integral $t$ of 0.45 meV (Table S2). The parallel spin alignment is observed along the stacking direction, with $J > 0$ (Table 2). Instead, intralayer magnetic interaction between adjacent paramagnetic centers is dominated by through-bond kinetic exchange with hopping integral $t$ of 14.2 meV. Indeed, neighboring atoms exhibit alternating spin alignments along the magnetic path, implying antiferromagnetic interactions within the layer, as confirmed by $J < 0$ (Table 2). The antiparallel spin alignment can be attributed to the super-exchange interaction between paramagnetic centers mediated by nonmagnetic Ph edge, as overwhelmingly observed in most metal oxides and radical COFs. The anisotropic magnetism and spin alignment in 2D COFs have been schematically illustrated in Figure 4d.

We would like to point out that, due to the difficulty in forming –C=C– linkage with Knoevenagel polycondensation reaction, the intralayer domain size is small in synthesized sp$^2$c-COF, while large domain size in the layer stacking direction is easy to achieve, so needle-like samples are formed as manifested by scanning electron microscope (see Figure S3 in Ref. 13). So, we attribute the ferromagnetic transition observed in iodine doped PPy-CC-Ph to the interlayer ferromagnetic interaction. We also believe that the intralayer antiferromagnetic interaction plays a prominent role in the formation of long-ranged magnetic order across the material, since one-dimensional magnetic structure is not stable. Besides, the kink in the experimental temperature dependence of spin susceptibility (see Figure S19 in Ref. 13) seems to indicate that antiferromagnetic interactions might exist and magnetic interactions in sp$^2$c-COF can be complicated, which need further experimental study.

In addition to PPy-CC-Ph, other three COFs with narrow valence band, PPy-CC-Naph, PPy-BO-Ph and PPy-CC-4FPh, exhibit similar magnetic behaviors, including fractional charge-transfer upon iodine doping and magnetic moment of almost one $\mu_B$ per node (Table 2). However, in PPy-BO-Ph with non-conjugated linkage, the in-plane magnetic coupling ($J$) is vanishingly small, which will reduce the three-dimensional magnetic structure to one-dimensional (1D). Due to the instability of 1D magnetic structure, long-ranged spin alignment in the out-of-plane direction may hardly develop in PPy-BO-Ph. It is also worth mentioning that the valence bands in PPy-CC-Naph and PPy-CC-4FPh are more localized than PPy-CC-Ph, which gives rise to more intense spin density centered on the pyrene node and stronger interlayer ferromagnetic interactions via direct potential exchange. Meanwhile, intralayer antiferromagnetic interactions in them are weaker due to the reduced intralayer hopping integral $t$ and smaller kinetic exchange ($4t^2/U$). So, we have attained two new prospective magnetic COFs, namely PPy-CC-Naph and PPy-CC-4FPh, with much larger interlayer ferromagnetic coupling $J$ (6.4 meV and 7.4 meV) than PPy-CC-Ph (1.7 meV), which indicates higher Curie temperature and better magnetic performance.

## CONCLUSIONS

To conclude, we proposed a promising approach to attain metal-free magnetism in nonmagnetic covalent organic frameworks by chemical doping at high dopant concentrations. Two requirements are to be met: the COFs should have narrow energy band and weak electronic couplings; the frontier orbital energies of the dopants should be close to those of the COFs. We also proposed three practical strategies to design COFs with narrow bands, which include choosing frontier orbital energy mismatched building blocks, applying non-conjugated linkages, and increasing structural distortion by introducing stereo-hindrance. The current work reveals the vital role of dopants in introducing paramagnetic centers to COFs. The dopants with frontier orbital energies close to COFs, such as p-dopants of I and Br and COFs whose valence band are localized on the pyrene node, not only inject charges to them but also flatten the energy bands and localize the electronic states via the formation of supramolecular charge-transfer complex. Based on these discoveries, we designed two new COFs, namely PPy-CC-Naph and PPy-CC-4FPh, and both of them show prospective magnetism after iodine doping. Finally, we unraveled the anisotropic magnetic interactions in these 2D COFs, where interlayer ferromagnetic and intralayer antiferromagnetic interactions are dominated by through-space potential exchange and through-bond kinetic exchange, respectively. We demonstrated that, since the magnetic coupling and conduction of opposing spin channels can be effectively modulated by manipulating the chemical building blocks, both metallic and half-metallic structures have been achieved. These findings not only shed light on how to develop new spin-polarized COFs, but also may pave the way for attaining magnetism in other nonmagnetic organic materials for applications in organic spintronic devices.

## METHODS

All the electronic structure calculations conducted in this research, including optimization of lattice parameters and atomic coordinates, were performed within the framework of density functional theory (DFT) as implemented in the Vienna ab-initio simulation package (VASP 5.3.5)[41] using the projector augmented wave (PAW)[42] method and Perdew−Burke−Ernzerhof (PBE)[43] exchange-correlation functional. London dispersion correction was further applied with Grimme's D3 approach,[44] and Heyd−Scuseria−Ernzerhof (HSE06)[45] hybrid functional was used to obtain accurate band gaps. The cut-off energy of 400 eV for plane-wave basis set was used in lattice and atomic coordinate optimizations, and 600 eV in static calculations. And the convergence criterion for forces on atoms during optimizations was set to 0.02 eVÅ$^{-1}$, while the energy convergence criterion in the self-consistent field iteration was set to $10^{-5}$ eV for optimizations and $10^{-6}$ eV for static calculations. The k-mesh of 1×1×5 and 1×1×3 was used in the optimization of pristine and doped COFs respectively, while denser k-mesh of 1×1×8 and 2×2×6 was used to obtain the converged charge density of optimized structures. The Bader charge and spin analyses were performed with the code implemented by Henkelman and co-workers.[46] To study the doping effect and injection of spins, a 1×1×2 supercell of COFs and two dopant atoms deposited into pores were constructed, so the ratio of the number of pyrene node to the dopant was 1:1 which resembles the experimental condition of high dopant concentration. The 1×1×4 and 2×2×2 supercells were built to study the interlayer and intralayer magnetic interactions, and k-mesh of 2×2×3 and 1×1×3 were used in the calculations, respectively. Spin polarization effect was considered in the calculation of doped COFs. The frontier orbital energies of building blocks of COFs and dopant atoms were obtained by the *Gaussian16* program package[47] using the long-range corrected functional ωB97XD and the def2tzvp basis set.

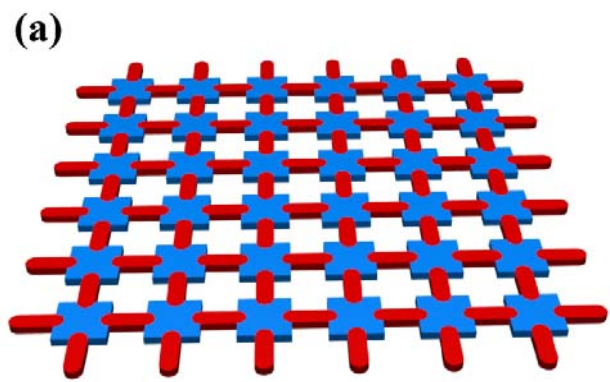

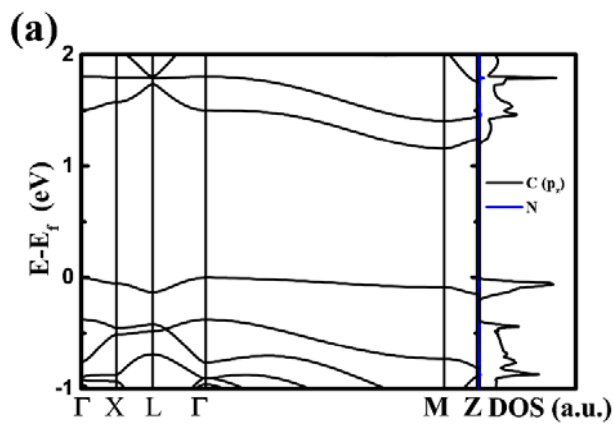

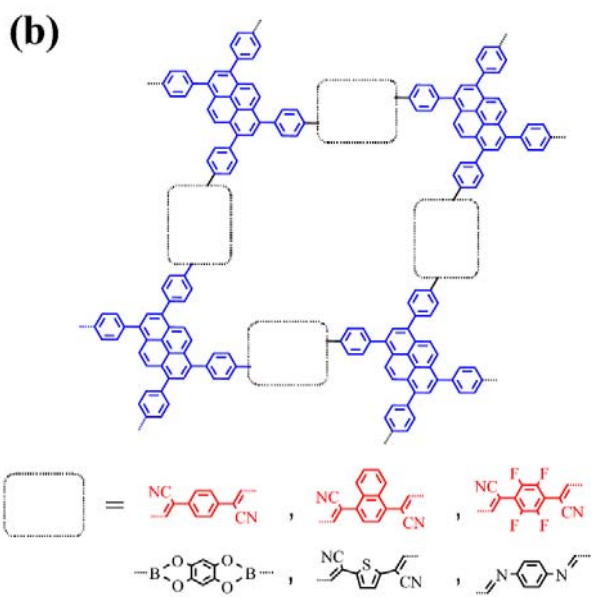

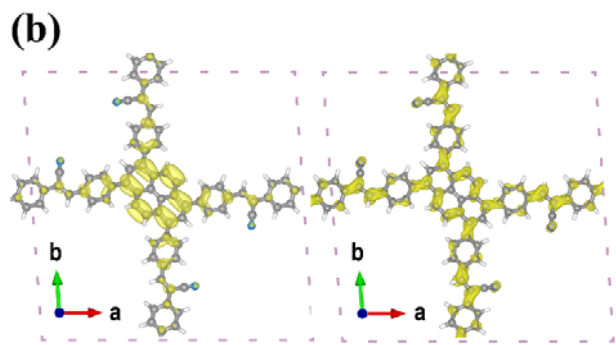

**Figure 1. Electronic structure of pristine PPh-CC-Ph. (a) Band structure and partial density of states. (b) Charge density at VBM (left) and CBM (right).**

**Scheme 1. (a) Topological structure and building blocks of 2D COFs with node (blue), edge and linkage (red) between them. (b) Chemical structure of PPy-CC-Ph, PPy-CC-Naph, PPy-CC-4FPh, PPy-BO-Ph, PPy-CC-1T and PPy-CN-Ph. PPy represents the pyrene node and it is short for tetrakis(4-formylphenyl)phenyl. Ph, Naph, 4FPh and 1T represent edges and are short for phenyl, naphthyl, 2,3,5,6-tetrafluorophenyl and thienyl respectively. CC, BO and CN represent C=C, boronate ester and C=N linkages.**

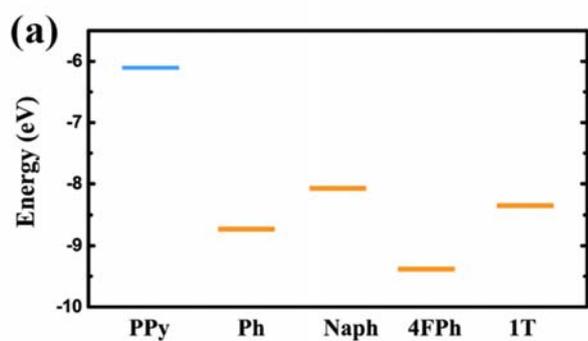

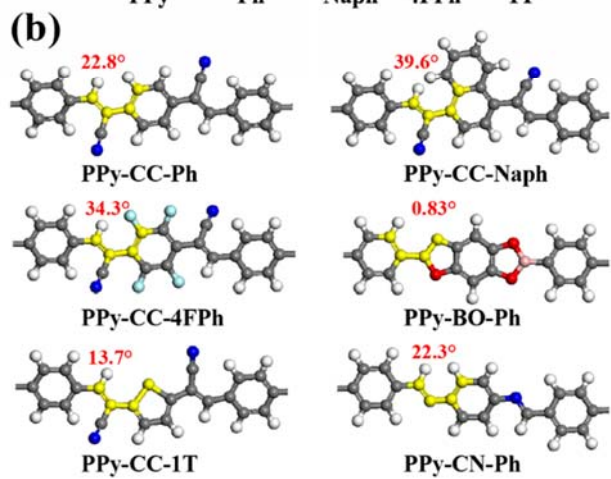

**Figure 2. (a) HOMO energy level diagram for building blocks of COFs. PPy represents the pyrene node, Ph, Naph, 4FPh and 1T represent edges of COFs, with linkages attached to them. (b) Structural distortion in COFs as indicated by the dihedral angle at the linkage. The four atoms constituting the dihedral angle are highlighted in yellow.**

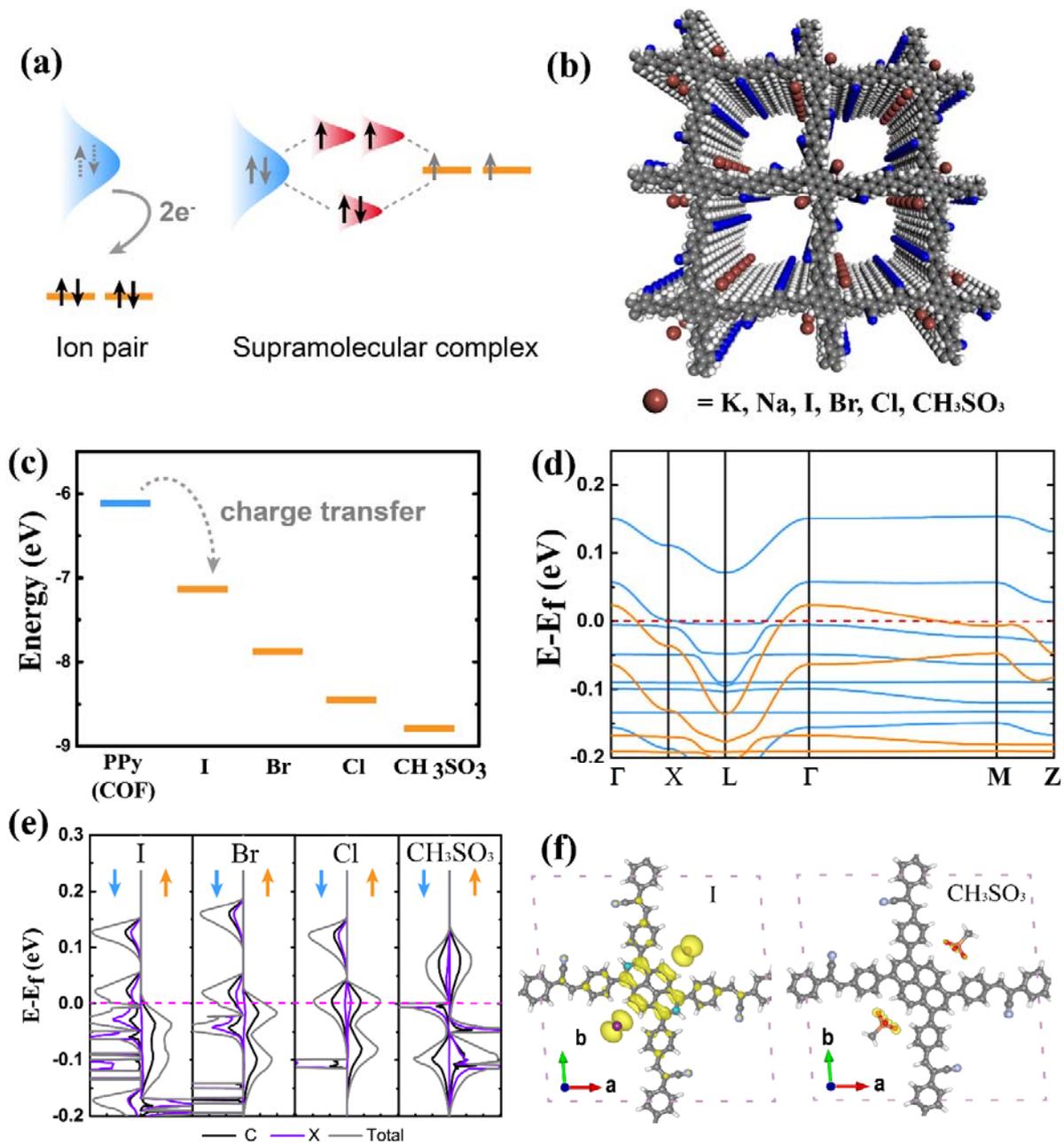

**Figure 3. Introducing unpaired spins to COFs by chemical doping.** (a) Two chemical doping mechanisms: ion-pair formation and supramolecular complex formation. (b) Structure of doped COFs. The ratio of pyrene node to the dopant is 1:1 and dopant atoms are deposited in pores every other layer. The color code of atoms is C: *grey*, H: *white*, N: *blue*, dopant: *brown*. (c) Frontier orbital energy level alignment of pyrene node in COFs and the p-dopants: I, Br, Cl and $CH_3SO_3$. (d) Spin-polarized band structure of I-doped PPy-CC-Ph. (e) Partial DOS of PPy-CC-Ph doped with I, Br, Cl and $CH_3SO_3$. X denotes the dopant. (f) Spin density distribution in I-doped and $CH_3SO_3$-doped PPy-CC-Ph.

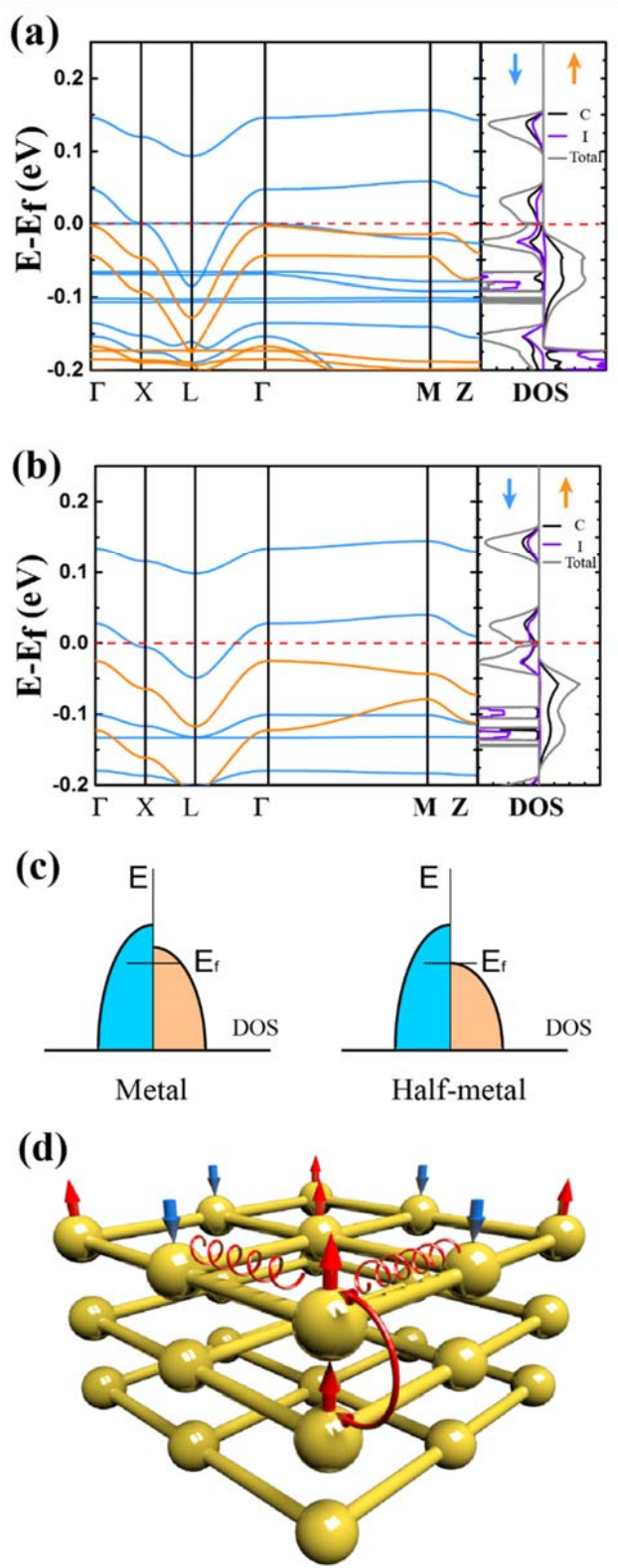

**Figure 4.** (a) Spin-polarized band structure and DOS of I-doped PPy-CC-Naph. (b) That of PPy-CC-4FPh. (c) Schematic plot of DOS for metallic and half-metallic spintronic materials. (d) Schematic illustration of anisotropic magnetic configurations in 2D COFs, with through-bond antiparallel and through-space parallel spin alignments for intralayer and interlayer interactions respectively.

**Table 1. Magnetic properties of PPy-CC-Ph doped with I, Br, Cl, CH$_3$SO$_3$, Na and K.**

|  | Dopant (X) | SOMO energy (eV) | Net charge on X | Magnetic moment ($\mu_B$ per node) | $U/W$ | pDOS on X (%) | $-\Delta E_P$ (meV)* |
|---|---|---|---|---|---|---|---|
| p-dopant | I | -7.12 | -0.49 | 0.98 | 1.08 | 20 | 45 |
|  | Br | -7.86 | -0.58 | 0.84 | 0.93 | 18 | 20 |
|  | Cl | -8.44 | -0.65 | 0.51 | 0.85 | 15 | 7.2 |
|  | CH$_3$SO$_3$ | -8.79 | -0.82 | 0.03 | 0.03 | 6 | 3.8 |
| n-dopant | Na |  | 0.9 | 0 | 0 | 0 | 0 |
|  | K |  | 0.9 | 0 | 0 | 0 | 0 |

*$\Delta E_P$ is the energy difference between spin-polarized and spin-unpolarized states.

**Table 2. Magnetic properties of PPy-CC-Ph, PPy-CC-Naph, PPy-CC-4FPh, PPy-BO-Ph, PPy-CC-1T and PPy-CN-Ph doped with I.**

| COFs | $W$ (eV)* | Net charge on I | Magnetic moment ($\mu_B$ per node) | $-\Delta E_P$ (meV)** | $U$ (meV) | $J$ (meV) | |
|---|---|---|---|---|---|---|---|
|  |  |  |  |  |  | Intralayer | Interlayer |
| PPy-CC-Ph | 0.23 | -0.49 | 0.98 | 45 | 124 | -1.8 | 1.7 |
| PPy-CC-Naph | 0.14 | -0.50 | 1.00 | 48 | 121 | -1.4 | 6.4 |
| PPy-CC-4FPh | 0.14 | -0.43 | 1.00 | 72 | 155 | -0.6 | 7.4 |
| PPy-BO-Ph | 0.20 | -0.43 | 0.98 | 69 | 128 | -0.1 | 2.0 |
| PPy-CC-1T | 0.42 | -0.49 | 0.65 | 47 | 46 |  |  |
| PPy-CN-Ph | 0.58 | -0.59 | 0.16 | 1.7 | 11 |  |  |

*Valence bandwidth of pristine COFs.

**$\Delta E_P$ is the energy difference between spin-polarized and spin-unpolarized states.

## ASSOCIATED CONTENT

**Supporting Information**

The Supporting Information is available free of charge on the ACS Publication website http://pubs.acs.org.

## AUTHOR INFORMATION

**Corresponding Author**

*E-mail: (D. W.) dong913@mail.tsinghua.edu.cn
**ORCID**
Dong Wang: 0000-0002-0594-0515
**Notes**
The authors declare no competing financial interest.

## ACKNOWLEDGMENT

This work is supported by the National Natural Science Foundation of China (Grant No. 21673123) and the Ministry of Science and Technology of China (Grant No. 2015CB655002). Computational resources are provided by the Tsinghua Supercomputing Center. The authors also acknowledge Dr. Jiajun Ren for his support on the calculation of electron affinity.